\documentclass[final]{elsarticle}
\usepackage{rotating}
\usepackage{hyperref}
\biboptions{authoryear}
\usepackage{amsmath}
\usepackage{amssymb}
\usepackage[T1]{fontenc}

\newcommand{\nlarge}{54}    %

\newcommand{\HI}{Hawai'i}
\newcommand{\halema}{Halema`uma`u}
\newcommand{\puo}{Pu`u `\={O}`\={o}}
\newcommand{\kila}{K\={\i}lauea}
\newcommand{\frange}[2]{\mbox{{#1}--{#2}~Hz}}
\newcommand{\refFig}[1]{Figure~\ref{#1}}
\newcommand{\refTab}[1]{Table~\ref{#1}}
\newcommand{\bN}{\mbox{${\bf N}$}}
\newcommand{\bS}{\mbox{${\bf S}$}}
\newcommand{\refFigab}[2]{Figures~\ref{#1}--\ref{#2}}
\begin{document}
\begin{frontmatter}
\title{Negative isotropic seismic moment tensors, migrating and cyclic seismicity during the 2018 summit collapse at \kila\ caldera}
\author[1]{Celso R. Alvizuri}
\author[2]{Robin S. Matoza}
\author[3]{Paul G. Okubo}
\address[1]{Institute of Earth Sciences, University of Lausanne}
\address[2]{Department of Earth Science and Earth Research Institute, University of California, Santa Barbara}
\address[3]{Department of Earth Sciences, University of \HI\ at Manoa, Honolulu, \HI}
\begin{abstract}
    The 2018 rift zone eruption of \kila\ volcano was accompanied by a remarkable and episodic collapse of its summit.
    Between May-August the eruption and collapse sequence included over 70,000 earthquakes (M$\geq$0) and 54 major earthquakes (M$\geq$5). 
    We analyzed the seismicity in the \kila\ summit region and estimated seismic full moment tensors with their uncertainties for the 54 M$\geq$5 events.
    These events occurred at almost daily intervals and were accompanied by intense seismicity which was concentrated between 0-3~km depths beneath the \halema\ pit crater.
    The hypocenters reveal partial elliptical patterns (map view) that migrated downward by $\sim$200~m.
    The moment tensors reveal remarkably consistent mechanisms, with negative isotropic source types and localized uncertainties, and vertical P-axis orientations.
    From the moment tensors we derived Poisson's ratios which are variable ($\nu=0.1-0.3)$ for the first half of the collapse events and converged to $\nu\sim0.28$ from June 26 onward.
\end{abstract}
\begin{keyword}
Seismic moment tensors,
non-double-couple,
isotropic,
seismicity,
caldera collapse,
Kilauea volcano.
\end{keyword}
\end{frontmatter}
\section{Introduction}
\label{sub:introduction_or_discussion}
On 2018 \kila\ volcano experienced its largest flank eruption in 200 years and a dramatic collapse of its caldera.
The eruption sequence included 
the largest earthquake on the island in 43 years (Mw6.9; 2018-05-04),
56 other large earthquakes (Mw$\ge$5) observed worldwide,
and over 70,000 earthquakes (Mw$\ge$0) across the island.
Most of the large earthquakes (\nlarge) occurred at depths 0-3~km beneath the summit region at \halema\ crater.

According to \cite{neal2019}, the 2018 flank eruption and caldera collapse at \kila\ are characterized by inflationary ground deformation starting mid-March,
rising lava lake levels at \puo\ and \halema\ craters through April, 
propagating seismicity and lava fountaning towards the Lower East Rift Zone (LERZ).
On May 1 the \halema\ summit began to deflate and lava lake levels began to drop;
on May 4 the Mw6.9 earthquake occurred at about 6~km beneath \kila's south flank.
Soon after the Mw6.9 earthquake, deflation of \kila's summit accelerated,
and by May~10 the lava lake level had dropped by more than 300~m.
Toward the end of May, the summit began to subside in episodic,
almost daily patterns,
and the crater floor dropping by several meters during each event.
These patterns are characterized by escalating earthquake swarms of up to 700 events per day,
each followed by a large earthquake (M$\ge$5) and a short period of nominal seismicity levels.
The last collapse event occurred on August 2, at about the same time as lava effusion stopped at the LERZ.

The 2018 activity at \kila\ prompted major field sampling and enhanced monitoring by the USGS Hawaiian Volcano Observatory and collaborators which, 
along with continuously operated networks,
offer unprecedented capabilities to observe and interpret sustained eruption and caldera collapse in \HI.
Since 1900 only six other caldera collapses have been documented in detail \citep{gudmundsson2016},
and seismological studies of caldera collapses include:
Piton de la Fournaise, R\'eunion Island (2007) \citep{michon2007},
Miyakejima, Japan (2000) \citep{geshi2002,Minson2007,Shuler2013data}, 
{B\'ardarbunga}, Iceland (2014-2015) \citep{nettles1998,gudmundsson2016}, 
and Fernandina, Gal\'apagos Islands (1968) \citep{simkin1970}.

In this study, we analyze the collapse events in the summit region at \halema\ crater by estimating seismic source mechanisms for the large events (Mw$\ge$5) and the intervening seismicity.
We estimate the full seismic moment tensors for the large events using seismic waveform data,
and use a comprehensive catalog of recently relocated hypocenters for the region.
Our moment tensor results reveal consistent collapse mechanisms as the collapse unfolds,
and the intervening seismicity shows migration of hypocenters.

\section{Data and Methods}

We estimate seismic moment tensors using waveforms from all possible broadband seismic stations within a 2,000 km radius,
available from IRIS-DMC. 
The waveforms were downloaded and processed using ObsPy, a python-based package for seismology \citep{obspy2010,megies2011obspy,obspy2015}.

The processing steps for each event were: 
(1) obtain three-component waveforms and metadata from IRIS-DMC; 
(2) remove instrument response using an 4-pole Butterworth filter with corner frequencies 0.005, 0.006, 10.0, and 15.0 Hz (flat bandpass \frange{0.006}{10.0});
(3) using the source-station azimuth and the sensor orientation angle, rotate horizontal components to radial and transverse directions.
Additional processing steps, such as cutting time windows and additional bandpass filtering were applied during the moment tensor inversions.
In our analysis, the waveforms at near stations (up to $\sim$15~km distance) show larger amplitude oscillations that may be related to near field effects,
while farther stations ($>$100~km) show waveforms with lower signal-to-noise ratio.
Therefore, in order to obtain robust moment tensor estimates we used data from stations within these distances.

We estimate full seismic moment tensors and their uncertainties for each event using the methodology described in \cite{Alvizuri2018-me}.
The method involves performing an efficient search over the full parameter space of moment tensors 
(lune longitude, lune latitude, strike, dip, and rake) including magnitude,
and uses a geometric parameterization for moment tensors and their uncertainty quantification
\citep[for more details and applications, see e.g.][]{AlvizuriTape2016,SilwalTape2016,Alvizuri2018-me}.
For each moment tensor in the parameter space,
synthetic seismograms are computed using a frequency-wavenumber approach \citep{ZhuRivera2002} with a 1D (layered) Earth model,
then the seismograms are compared with observed waveforms via a misfit function.
The synthetic seismograms for this study were computed using a 1D layered model for the region obtained from CRUST1.0 \citep{Crust1}.

Our moment tensor methodology has proven useful for earthquake source characterization for a range of settings in the Earth,
including Uturuncu volcano in southwest Bolivia \citep{AlvizuriTape2016-me}; 
tectonic events at a subduction zone in southcentral Alaska \citep{Alvizuri2018-me};
events possibly related to metamorphism in the Himalaya lower crust \citep{Alvizuri2018c-me};
and nuclear tests and cavity collapses in western USA and in North Korea \citep{Alvizuri2018-me,Alvizuri2018b-me}.

We complemented our moment tensor results by analyzing the temporal and spatial distribution of hypocenters at \kila\ volcano for 2018.
The hypocenters were obtained from a recent comprehensive study which relocated seismicity from 32 years (1986-2018) in the Island of \HI\ \citep{matoza2020relocations}.
The relocations were done with the GrowClust algorithm which combines waveform cross-correlation, 
hierarchical cluster analysis and relative relocations;
for details, see \cite{matoza2013} and \cite{matoza2020relocations}.
We focused on the seismicity beneath the summit at \halema\ crater, %
between longitudes [-155.30, -155.24], 
latitudes [19.38, 19.44],
and depths [0, 3]~km.
In total \nlarge\ events above magnitude Mw~5 occurred in the \kila\ summit region over a period of three months.

\section{Results}

\subsection{Seismicity}%
\begin{figure}[ht]
    \centering
\includegraphics[width=14cm]{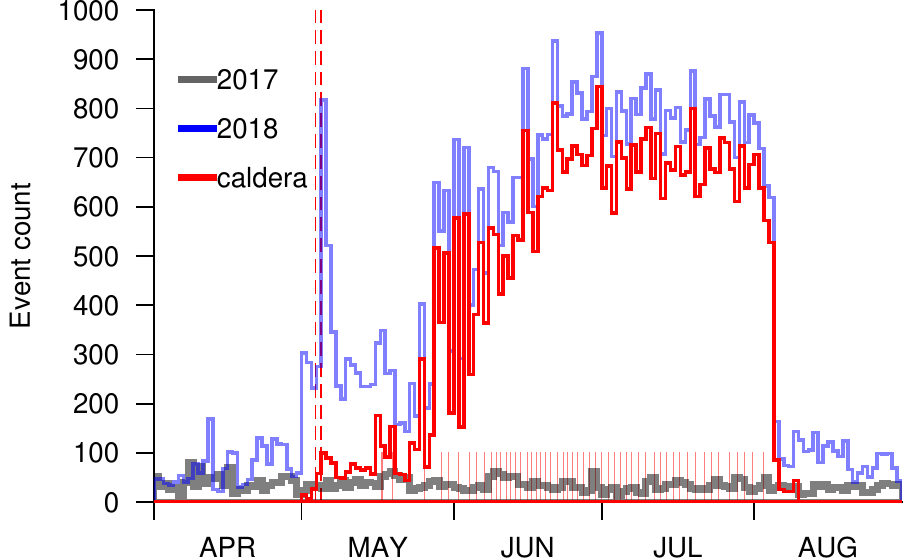} 
    \caption{Seismicity levels for \kila\ (red lines) and the greater region (blue).
    The daily seismicity at \kila\ from 2017 (gray lines) up to April 2018 remained at mostly background levels of about 30 events per day.
    On 1 May 2018 the number of events increased suddenly to about 300/day, followed by another increase to about 800/day after the Mw 6.9 event on 2018-05-04 (red dashed line).
    On about May 20 the number of events increased again and reached levels of about 800/day.
    This rate is sustained until August 6 when the seismicity dropped close to background levels.
    In total the seismicity increased by two orders of magnitude from background levels to the East Rift Zone eruption.
    The 54 seismic events at the caldera occurred at regular intervals.
    The short vertical lines (red) show the times of \nlarge\ seismic events.
    }
    \label{fig:seismicity_histograms}
\end{figure}

\begin{figure}[ht]
    \centering
    \includegraphics[width=14cm]{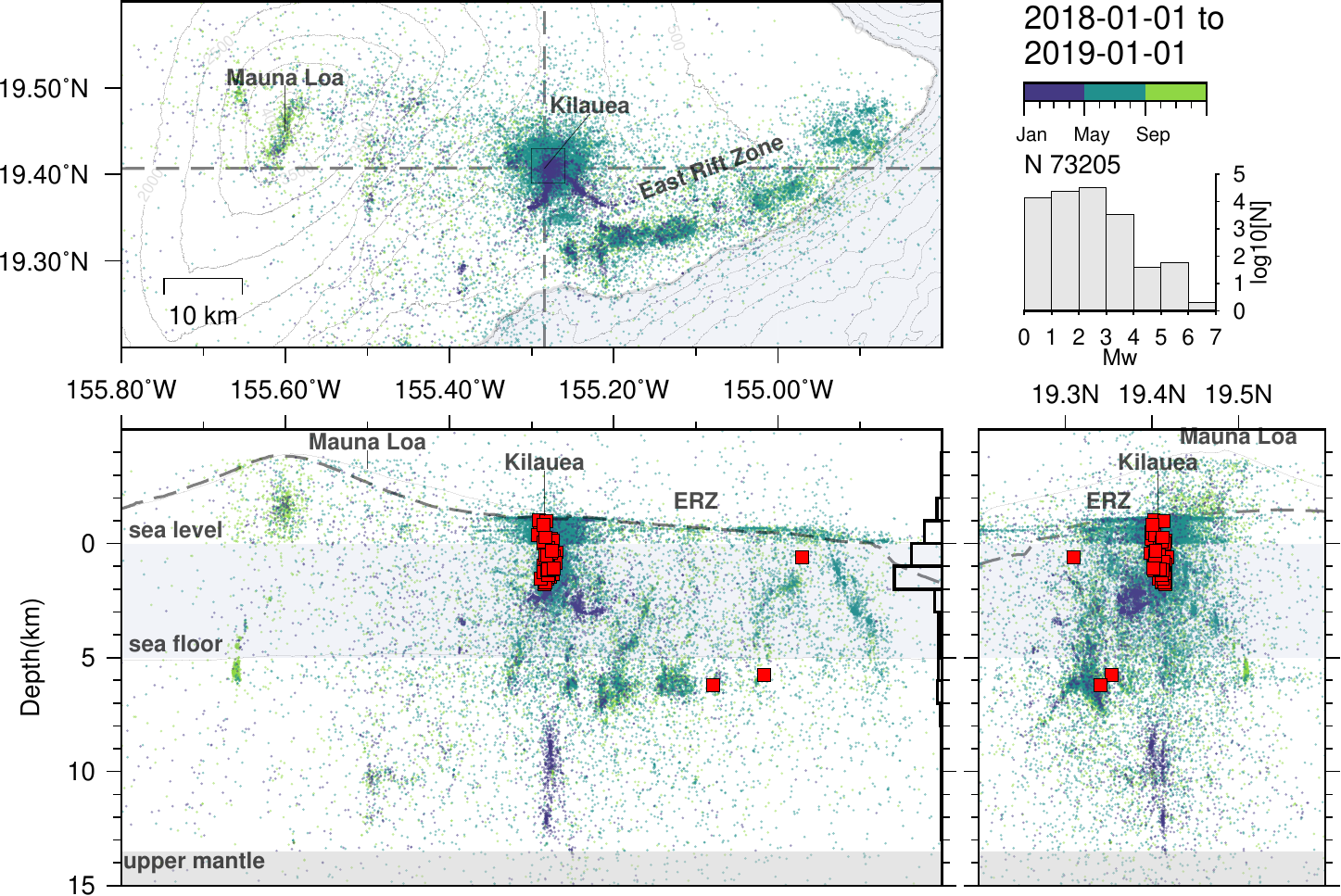}
    \caption{
    The 2018 seismicity at \kila\ in map view and cross sections (vertical exaggeration is 3X; the crosshairs are centered at Halema'uma'u crater).
    Each seismic event is colored by the time it occurred (early in the year = purple, late in the year = green).
    The red squares are events of magnitudes Mw$\ge$5, most of which occurred beneath \halema.
    Between Nov. 2017-Mar.~2018, there is an increase in seismicity beneath \halema\ at depths 7-13~km.
    In the months following, the seismicity becomes shallower ($<$5~km) and concentrates in \halema\ and the East Rift Zone.
    Between September-December there is a slight increase in seismicity beneath Mauna Loa.
    }
    \label{fig:seismicity_xsect_full}
\end{figure}

\label{subsub:seismicity}
The seismicity at \kila\ shows three main periods during 2018
(\refFig{fig:seismicity_histograms}):
(1) January-April, seismic activity at background levels of about 30 events per day;
(2) May-August,    activity changes abruptly to 300/day and again to sustained levels up to 800/day;
(3) August 4-December,   activity returns to background levels.
\refFig{fig:seismicity_xsect_full}
shows hypocenters on map and cross sections for the three periods above.

During the first period (Jan-May) the seismicity was concentrated beneath the \halema\ summit region
and was typical for the region (with approximately the same distribution from 1986-2017).
The seismicity occurred rom the surface to about 3~km below sea level (bsl) and from 7-13~km bsl,
and in the upper West and upper East Rift Zones from 1-3~km bsl.
The gap between $\sim$3-7~km is attributed to the relatively aseismic magma storage reservoir.
The next period (May-Aug) shows elevated seismicity in the same \halema\ summit region
and along ERZ (depths 3-7~km bsl).
In the next period (Aug-Dec) the seismic activity decreased to background levels (though slightly more elevated than 2017 levels),
and increased beneath Mauna Loa summit.

\begin{figure}[ht]
    \centering
    \begin{tabular}{c}
        \includegraphics[width=13cm]{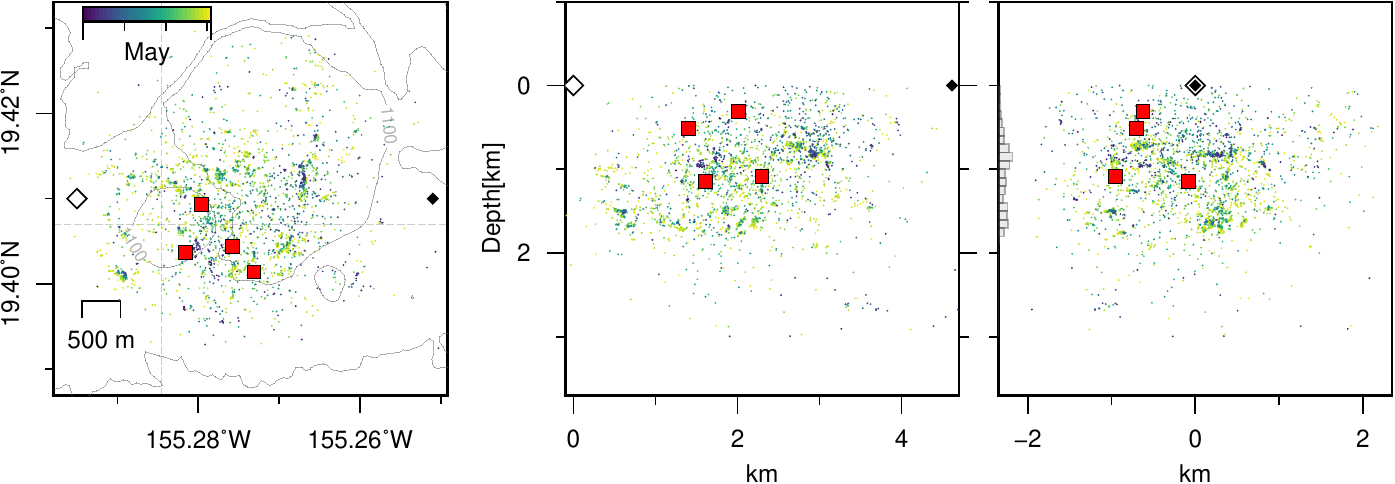} \\ 
        \includegraphics[width=13cm]{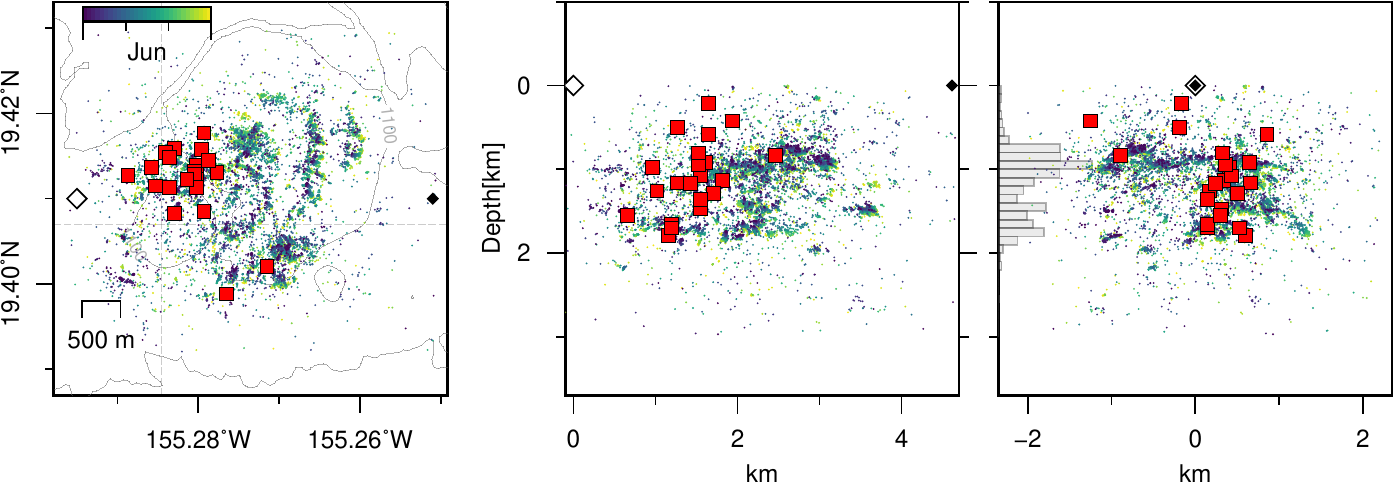} \\ 
        \includegraphics[width=13cm]{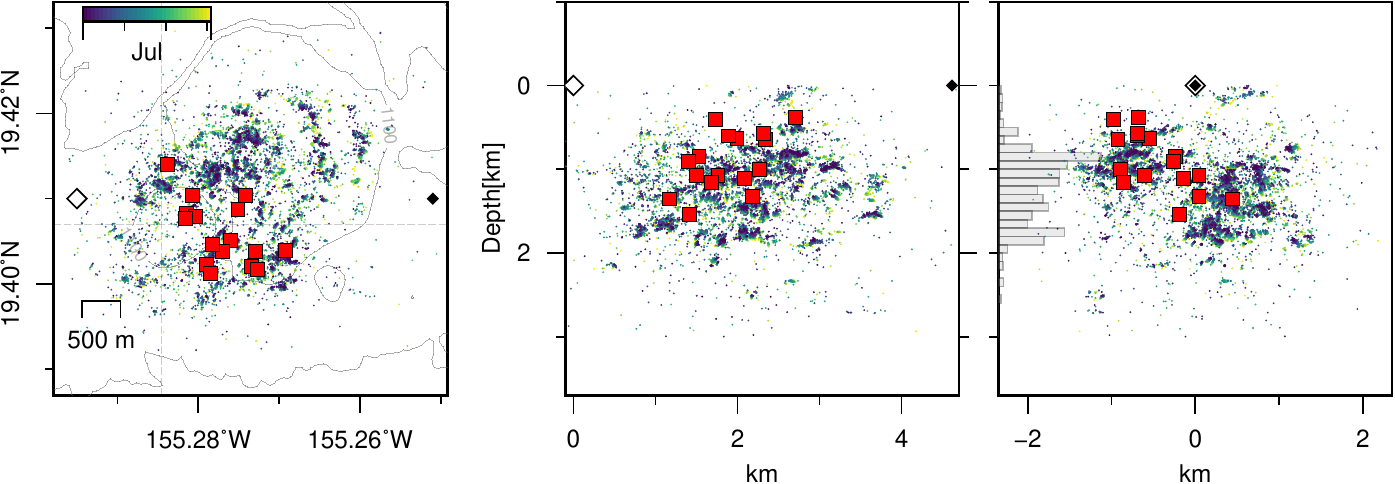} \\ 
        \includegraphics[width=13cm]{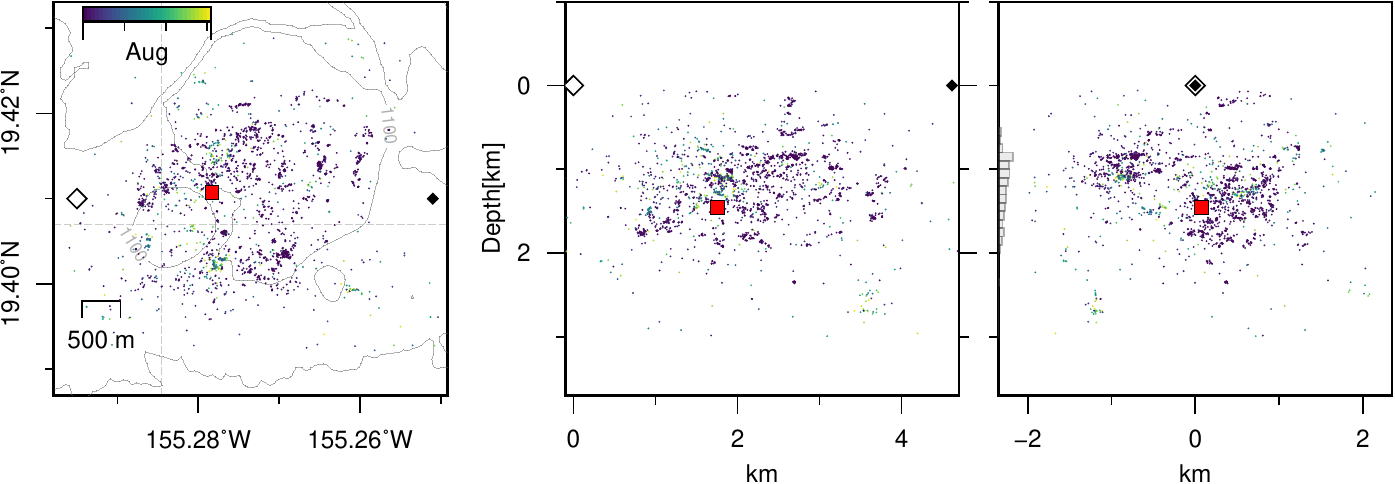} \\ 
    \end{tabular}
    \caption{
        Views into four months of seismicity at \kila\ from May-August, 2018.
        Each row shows one month of hypocenters (top row is for May, bottom row is for August).
        The left column shows map views, the center and right columns show cross-sections looking toward north and toward west.
        By June-July the relocated seismicity highlights prominent structures including elliptical patterns, which are less visible earlier (May) with fewer cumulative events.
        During these months the seismicity also migrated radially outward (map view) and downward (cross-section view).
    }
    \label{fig:seismicity_view_kila}
\end{figure}

\refFig{fig:seismicity_view_kila} shows a closer view of the seismicity at \halema\ during May-August (panels a-d).
Seismic activity in May increases and is generally diffuse,
then changes into distinct (partial) elliptical patterns throughout June-July,
then changes back into diffuse in August until it abruptly drops to background levels on 3 August.

\begin{figure}[ht]
    \centering
    \begin{tabular}{cc}
    \includegraphics[width=7cm]{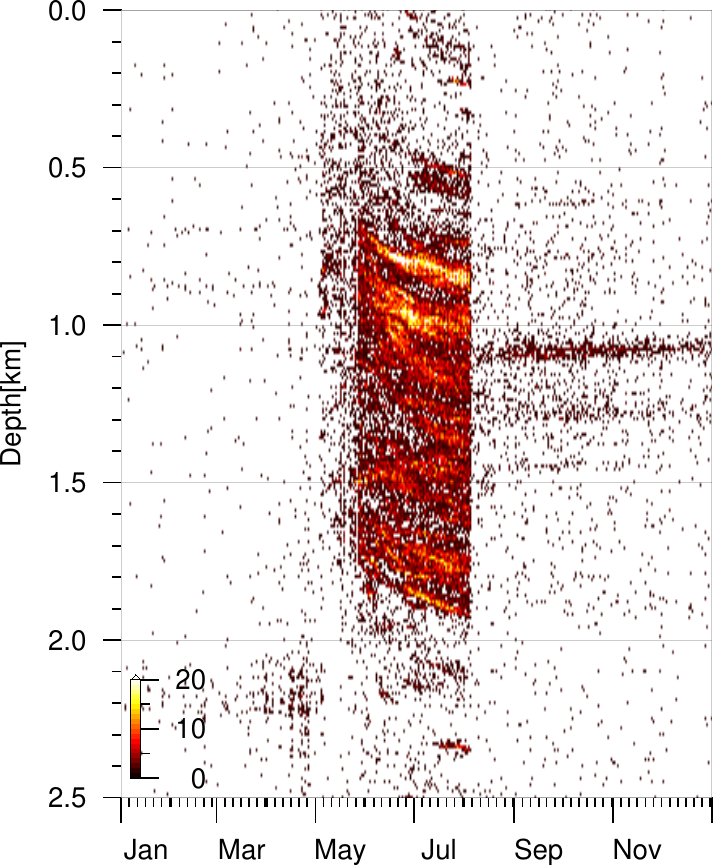}&
    \includegraphics[width=7cm]{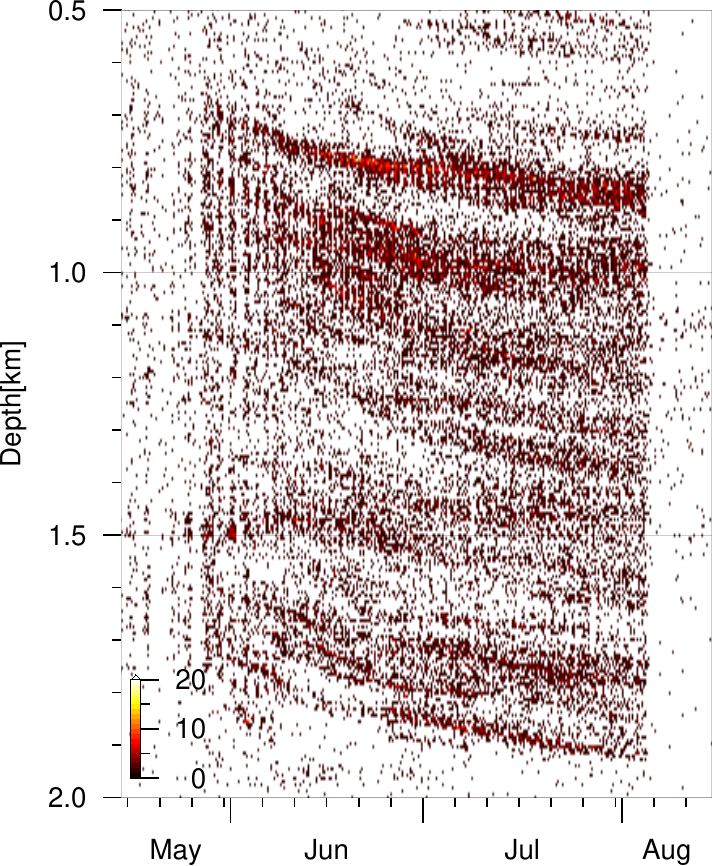}
    \end{tabular}
    \caption{
        Hypocenter count by depth-time (heatmap) at \kila\ during 2018.
        The heatmaps were computed using hypocenters within the same boundaries as in \refFig{fig:seismicity_view_kila}, at 1-day bins and 10-meter depth intervals for: (a) the entire year and (b) the period May-August.
        During May the seismicity increased to sustained levels (above 20 events per bin).
        During May the seismicity increased to sustained levels (ab        From late May until early August the peaks in seismicity are concentrated between depths 0.5-2 km.
        During this period the seismic activity migrated downward by about 200 m.
    }
    \label{fig:heatmap_seismicity}
\end{figure}

During this period the hypocenters migrated radially outward (map view) and downward (cross-section view);
\refFig{fig:heatmap_seismicity} shows hypocentral depths with time beneath the \halema\ summit region for 2018.
This result shows that the peaks in seismicity are concentrated between depths 0.5-2 km beneath the summit,
and throughout June-July they migrate downward by about 200~m.

\subsection{Seismic moment tensors}%
\label{sub:seismic_moment_tensors}

In our results the synthetic seismograms show good agreement with observations at most seismic stations.
The moment tensor uncertainty estimates for each event show best fitting mechanisms that are localized toward the negative isotropic ($-$ISO) region on the lune (see Supplement);
for details on the uncertainty analysis, see \cite{Alvizuri2018-me}.
The supplement shows waveform fits and uncertainty estimates for the \nlarge\ events analyzed here.

\begin{figure}[ht]
    \centering
    \begin{tabular}{cc}
        \includegraphics[width=0.50\linewidth]{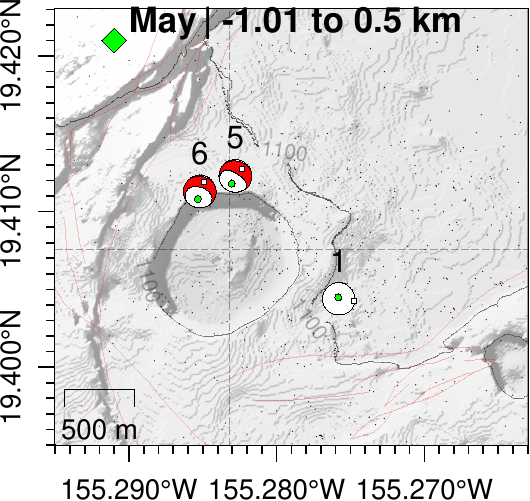} & 
        \includegraphics[width=0.50\linewidth]{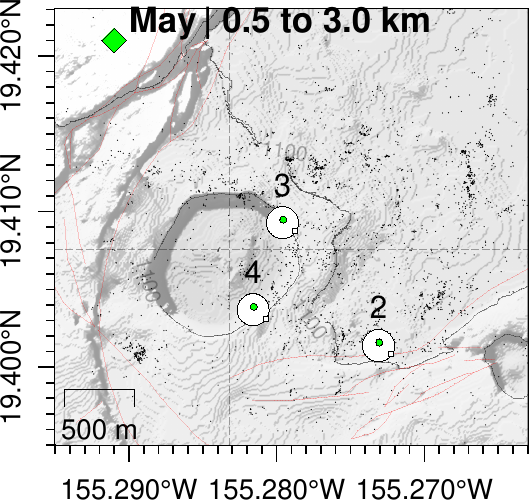} \\ 
        \includegraphics[width=0.50\linewidth]{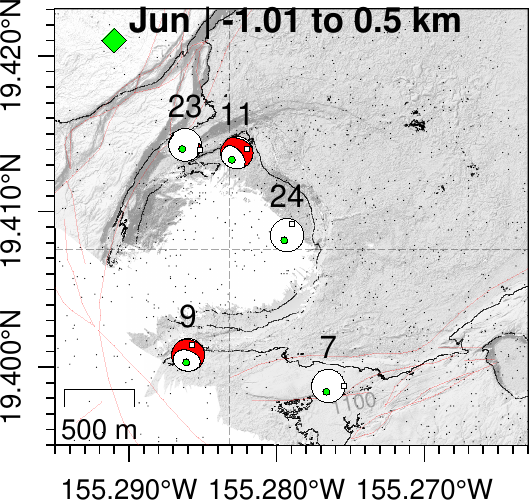} & 
        \includegraphics[width=0.50\linewidth]{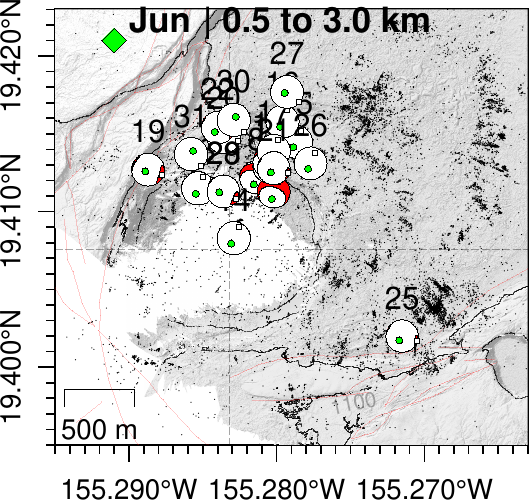} \\ 
        \includegraphics[width=0.50\linewidth]{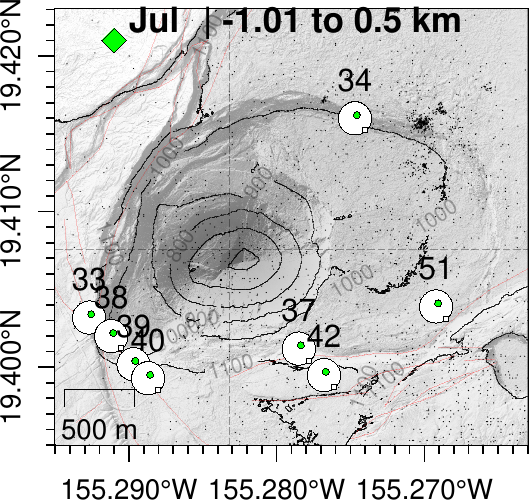} & 
        \includegraphics[width=0.50\linewidth]{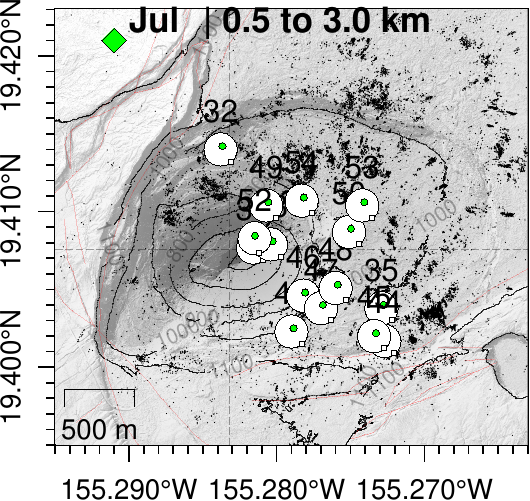} \\ 
    \end{tabular}
    \caption{
        Closer view into the seismicity and moment tensors at \halema\ between May-July.
        The background elevation model for May is from ETOPO1, and for June-July from two LiDAR campaigns by the USGS.
        The green diamond shows tiltmeter station UWD.
        During this period the summit at \halema\ crater caved downward by up to 500~m while the crater rim expanded by about 1000~m.
        Between May-July several of the large events (depths $-$1 to 0.5~km bsl; left column) occur close to the crater rim.
        At depths $0.5-3$~km (bsl; right column) the mechanisms cluster (June) to the north of the previous crater, and (July) to the east.
}
    \label{fig:map_kilauea2018_caldera_bb}
\end{figure}

The best-fitting moment tensors for the \nlarge\ events show (\refTab{tab:table1}) consistent mechanisms 
with their P-axes oriented vertically and magnitudes between Mw4.9--5.6.
The source durations for the first 5/\nlarge\ events require source durations that decrease from 20 to 5 seconds,
while the remaining 49/\nlarge\ events range between 1-2 seconds (\refTab{tab:table1}).
\refFig{fig:map_kilauea2018_caldera_bb} shows epicenters and moment tensors at \halema\, together with elevation models during that period.

\begin{figure}[ht]
    \centering
    \includegraphics[width=16cm]{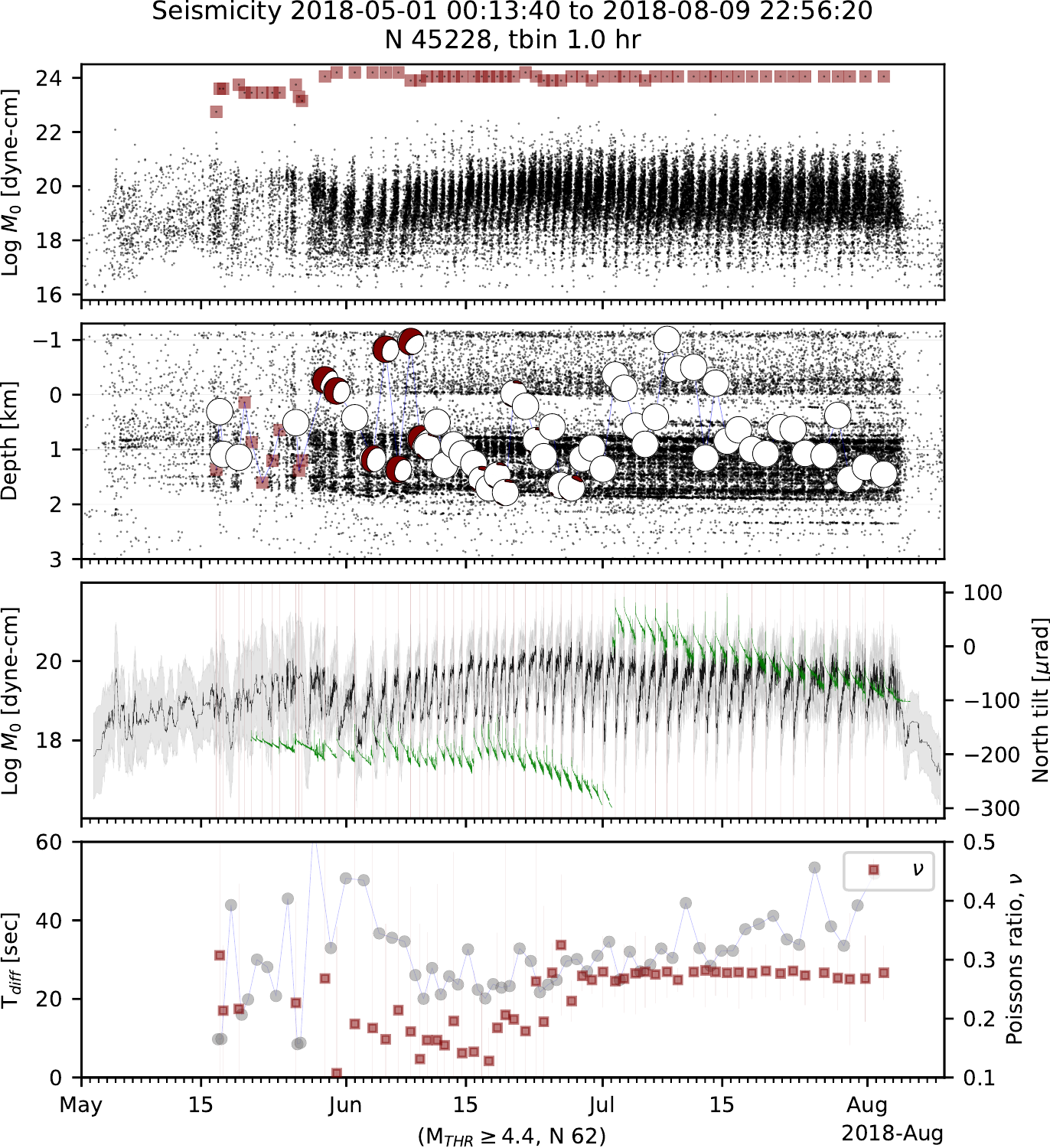}
    \caption{
        Results from the caldera collapse, period May 1-August 10:
        (a) seismic moments from all the events in \refFig{fig:seismicity_view_kila}; red squares highlight the large events (Mw$\geq$5) analyzed here;
        (b) hypocenter depths and focal mechanisms for the large events;
        (c) median seismic moment (black lines), mean absolute deviation (gray), times of the large events (red), and tiltmeter data (green) for station UWD (north component; the offset is due to recalibration of the sensor);
        (d) time difference (T$_{\text{diff}}$; gray circles) between consecutive events (M$_{\text{THR}}\geq4.4$) and Poisson's ratios (red) estimated from the moment tensors.
        (Below the magnitude threshold M$_{\text{THR}}=4.4$, T$_{\text{diff}}$ approaches zero and may indicate the event populations are not causally related \cite[e.g.][]{gardner1974}.
        For the mechanisms that we analyzed, T$_{\text{diff}}$ shows a trend between 20-40 seconds.
        Similar M$_{\text{THR}}$ values were also observed in Miyakejima, Piton de la Fournaise and Fernandina \citep[e.g.][and references there]{michon2011}).
    }
    \label{fig:plot_depth_time_seismicity_bb}
\end{figure}

The seismic moment tensor can also be related with a single-process source model introduced by \cite{AkiRichardsE1}
(for details and applications to other settings, see e.g. \cite{DufumierRivera1997,TapeTape2013,AlvizuriTape2016,Alvizuri2018c-me}).
In this model the source is represented by (possibly oblique) slip on a planar fault,
and is characterized by an angle $\alpha$ between normal $\bN$ and slip $\bS$ vectors,
and Poisson's ratio $\nu$.
For the \kila\ events up to June~25 their Poisson's ratios range between $\nu=0.1-0.3$;
from June~26-on they settle on consistent values near $\nu=0.28$ 
(\refFig{fig:plot_depth_time_seismicity_bb}d).
 
\refFig{fig:plot_depth_time_seismicity_bb} summarizes several observations and results with time:
(a) seismic moment; 
(b) hypocentral depths;
(c) median seismic moment and times of the major seismic events, tilt from station UWD (north component) near the northwest caldera rim;
(d) time difference between consecutive events, Poisson's ratios estimated from the moment tensors.

\section{Discussion}%
\label{sec:discussion}

Our moment tensor results for the M$\ge$5 events at \kila\ reveal consistent mechanisms
with mainly vertical P-axes orientations,
source types that cluster toward $-$ISO with (Figure~S3),
and source durations that decrease from 20 to 5~sec and converge to 2~sec from May~30-onwards
(\refFigab{fig:plot_depth_time_seismicity_bb}{fig:map_kilauea2018_caldera_bb}).

Similar mechanisms were also observed from the caldera collapse at Miyakejima \citep{Shuler2013models}.
In both settings the main seismic events take place after the start of volcanic eruption.
The moment tensors for Miyakejima show CLVD mechanisms with mainly vertical P axes
(though their analysis assumes deviatoric moment tensors)
and source durations in the order of 50-60~seconds.

Seismic source studies at volcanoes may combine moment tensors with single forces to study
various processes such as fluid-rock interaction \citep{kumagai2005,chouet2010,matoza2015},
and caldera collapse events \citep[e.g.,][]{kumagai2001,duputel2019}.
The events we analyzed for \kila\ may also arise from combinations of similar processes,
though the waveforms in our results show the moment tensor alone adequately fits the observations.

Non-double-couple moment tensors may also arise as artifacts from
imperfect Earth models, anisotropy, curved faults, etc. \citep[e.g.,][]{Kawasaki1981,Frohlich1994,Julian1998a}.
These are known tradeoffs in full moment tensor estimation,
and can also be addressed with 
multiple force systems, finite source studies \citep[e.g.,][]{fichtner2010drop}, more accurate structure models, etc.

Seismic moment-tensor studies for nuclear explosions at the Nevada Test Site (NTS; USA) and Punggye-Ri (North Korea)
reveal mechanisms with $+$ISO parameters for the explosions, 
and secondary events following some explosions show $-$ISO parameters and are presumed cavity collapses
\citep[e.g.,][]{Ford2009,Chiang2014,Cesca2017,Alvizuri2018-me,Alvizuri2018b-me}.

Field studies at NTS \citep[e.g.,][]{houser1969,Masse1981} %
and at quarry sites \citep[e.g.,][]{scandone1990},
and analogue sandbox experiments \citep{acocella2007,ruch2012} 
provide insight into the kinematic evolution of caldera collapses from small (cm) to intermediate scales (100s m).
They show block-like collapse structures and fault systems that develop near the surface,
with a range of radial and concentric cracks
and chimney collapse formations at depth.

A recent study of the collapse events at \halema\ in 2018 used GPS, tiltmeter and aerial observations,
and characterized the summit collapse as a funnel-like geometry,
with piston-like slumping of coherent blocks, and block areas on the order of 1.5~km$^2$ \citep{anderson2019}.

In our analysis the hypocenters beneath \halema\ (including the M$\geq$5 events) 
were concentrated between 0-3~km depths,
and between May-August migrated downward by about 200~m.
During this period the summit crater caved downward by up to 500~m while the crater rim expanded by about 1000~m.
The epicenters in some areas form partial elliptical patterns and radial streaks (depths 0.5$\sim$2.0~km),
and some large, shallower events (0$\sim$1.5 km depths) approximately follow the expanding contours of the crater
(\refFig{fig:map_kilauea2018_caldera_bb}).

During May-August the seismicity beneath the summit also shows almost daily cycles of earthquakes ramping up in count and moment release 
(similar patterns were also observed by \cite{butler2020}).
Each seismicity cycle was followed by a large event (M$\geq$5) and approximately 1~hr of relatively quiet periods (N$\leq$10).
This style in cyclic seismicity continued until $\sim$Jun 28, %
where it changed into more sudden onsets;
all cycles were still followed by a large event and a 1~hr quiet period.
(The supplement shows additional views and shorter time intervals).

The seismicity cycles also coincide with tiltmeter observations from station UWD (\refFig{fig:plot_depth_time_seismicity_bb}) to the west of the caldera rim,
which shows a long period trend of tilt to the southeast, towards \halema\
and intervening and periodic offsets that coincide with the times of the large earthquakes, and point away from \halema.
A recent study \citep{segall2019} suggests that these trends follow 
a primarily deflation process (revealed by subsidence at a GPS station at the \halema\ rim)
and a secondary process of ash emissions as observed by the radially outward transients (up to 89~$\mu$rad at UWD).
Similar tiltmeter transients were also observed on 2017 at Piton de la Fournaise (PdF),
and may reflect a continuum of deformation from the roof of the magma chamber to the surface \citep{michon2007,michon2009}.

The episodic collapses at \kila\ also appear coupled to pressure changes within the magmatic system.
A recent study estimated lava effusion rates at a newly developed fissure %
in the LERZ using ground-based video and time-lapsed images \citep{patrick2019},
and they observe dual cycles in lava eruption rates,
one with periods of 5-10 minutes where effusion rates change from $\sim$~350--1750m$^3$s$^{-1}$;
another as long-term surges in effusion rates and occurring no later than 20~min after the collapse events,
with effusion rates changing from $\sim$~300--500m$^3$s$^{-1}$ before collapse,
up to $\sim$~1400m$^3$s$^{-1}$ after collapse. %
This suggests a hydraulic connection where the summit reservoir provides pressurized magma to the flank vents,
in turn the flank vents regulate draining at the summit reservoir,
and where the summit collapse events induce pressure surges within the magma conduits.

The time difference T$_\text{DIFF}$ between consecutive events at \kila\ (with M$>4.4$) vary between 10-60~sec until June~9,
and from June~10-on they converge on a trend that increases from about 20-40~sec.
Other timings based on tiltmeter data were previously observed to range from 50-100~sec at PdF to 500-1500~sec at Miyakejima \citep{michon2011},
and together with a piston intrusion model \citep{kumagai2001},
they relate to frictional difference $F_{SD}$, piston geometry and mass, and lava effusion rates.
These were variable for \kila\ \citep[e.g.,][]{anderson2019,patrick2019}.
Nevertheless assuming all other variables constant, then $F_{SD} \propto \text{T}_\text{DIFF}$.
Our results show that T$_\text{DIFF}$ decreases from T$_\text{DIFF}=30-5$~sec. between May~15-30 and converges to 2~sec. starting June~9.

Core samples down to $\sim$1200~m beneath \halema\ reveal mainly basalt \citep{zablocki1974,keller1979},
and a seismic tomography study for \kila\ estimated Poisson's ratios between $\nu=0.25-0.32$ \citep{dawson1999}. %
Lab experiments where basalt samples are subjected to cyclic loading show incremental Poisson's ratios after each cycle,
up to $\nu=0.3$ \citep[e.g.,][]{schultz1993},
and $\nu=0.5$ for samples from Mt. Etna \citep{heap2009}.
Our Poisson's ratio estimates for \kila\ are initially variable but converge ($\nu\sim 0.28$) after June~26 and may reflect similar loading cycles.

The shallow magma reservoir beneath \halema\ is estimated to be a complex system 
comprising a plexus of sills, dikes, and magma filled cracks 
\citep{fiske1969,dawson1999,chouet2010}.
Towards the end of May the lava lake at \halema\ crater was no longer visible, 
which may also indicate partial vacating from the magma reservoir.
The Poisson's ratios in our results trend from $\nu=0.3$ to 0.1 between May until June~18,
and may reflect staggered and deepening collapses within the plexus that became evacuated.
The consistency in the moment tensor mechanisms and $\nu$ estimates from June~19-onwards may reflect 
collapses within a setting that is more consolidated and less accomodating.
\section{Conclusion}%
\label{sec:Conclusion}
The 2018 caldera collapse at \kila\ volcano was accompanied by 
\nlarge\ major (Mw$\ge$5) earthquakes and intervening seismicity which
was concentrated at depths 0-3~km beneath the summit area at \halema\ pit crater.
The intervening seismicity migrated downward by $\sim$200~m, approximately in daily cycles.
The seismicity in some locations formed partial elliptical patterns (map view),
and patterns radially outward from the pit crater (view at depth).
We estimated seismic full moment tensors for the \nlarge\ (Mw$\ge$5) events using waveform data from broadband seismic sensors.
The results reveal moment tensors with consistent negative isotropic mechanisms and vertical P-axes orientations.
Poisson's ratios estimated from the moment tensors are initially variable ($\nu=0.1-0.3$) and converge to $\nu\sim 0.28$ from June~26-onwards.
Incremental Poisson's ratios for the collapse events may reflect loading cycles that are observed in lab experiments.
The initial variability and later consistency in $\nu$ estimates may reflect
conditions within the roof blocks and plexus that are at first accomodating and variable,
and later become consolidated.
The negative isotropic mechanisms may reflect collapses within the evacuated plexus that comprise the magma chamber.

\clearpage\pagebreak
\begin{figure}[ht]
    \centering
    \includegraphics[width=8cm]{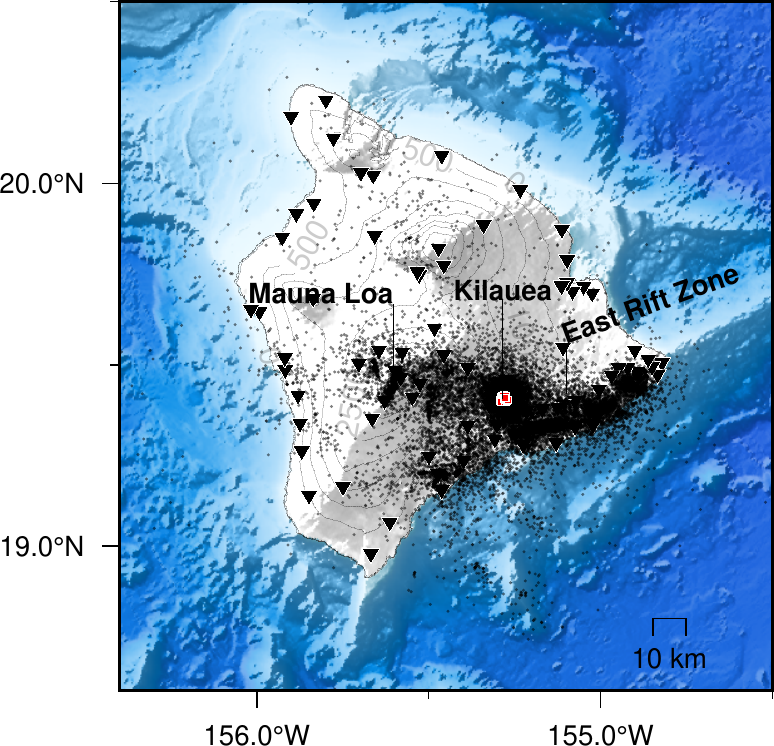}
    \caption{
        Epicenters (black dots) on the Island of \HI\ in 2018 based on a recent catalog of relocated hypocenters \citep{matoza2020relocations}.
        Most of the seismicity for the year was generated during the three months of eruption at the Lower East Rift Zone and during the summit collapse at \kila\ caldera.
        In this study we focus on all seismic events with magnitudes M$\ge$5 (\nlarge\ total; red squares) which occurred near the summit at \halema\ pit crater, and estimated their seismic full moment tensors using all possible waveform data from broadband seismic stations (black triangles).
        We also analyze the spatial and temporal distribution of the intervening seismicity (over 70,000 events).
}
    \label{fig:stamap}
\end{figure}

\clearpage\pagebreak

\clearpage\pagebreak

\clearpage\pagebreak

\clearpage\pagebreak

\clearpage\pagebreak

\clearpage\pagebreak

\clearpage\pagebreak
  \begin{sidewaystable}
    \caption{Moment tensor estimates for the M$\geq$5 events at \kila.
    Origin times and hypocenters are from \cite{matoza2020relocations}.
    Depths are with respect to sea level; negative up.
    The moment tensor estimates were computed at discrete 1-km depths beneath the summit surface; negative depths were set to 1~km depth.
    The moment tensors are parameterized using the lune longitude and latitude, strike, dip, slip and magnitude.
    VR (\%) measures the waveform fit between observed and synthetic seismograms.
    }
    \label{tab:table1}
    \footnotesize
    \begin{tabular}{rccccccccccccc}
        \hline
        No. & Origin time & Lon         & Lat           & Dep & Inv & Mw & Dur. & Lune             & Lune              & Strike        & Dip           & Slip          & VR \\
            &    (UTC)    &$(^{\circ})$ &$(^{\circ})$   &(km) & (km)&    & (sec)    & Lon $(^{\circ})$ & Lat $(^{\circ})$  & $(^{\circ})$  & $(^{\circ})$  & $(^{\circ})$  &    \\
        \hline
 1 & 2018-05-17T04:15:30.0459 &  -155.2758 &    19.4044 &   0.308 & 1 & 5.00 & 20    &    24 &   -43 &  58 &  61 &  -56 & 56 \\
 2 & 2018-05-17T14:04:11.1334 &  -155.2731 &    19.4014 &   1.088 & 1 & 5.10 & 15    &    16 &   -53 &  35 &  50 &  -88 & 86 \\
 3 & 2018-05-19T09:58:33.3367 &  -155.2796 &    19.4093 &   1.142 & 1 & 4.90 & 10    &    20 &   -43 &  62 &  63 &  -69 & 77 \\
 4 & 2018-05-26T02:15:52.2337 &  -155.2816 &    19.4037 &   0.510 & 1 & 5.00 & 15    &    18 &   -44 &  61 &  60 &  -72 & 76 \\
 5 & 2018-05-29T11:56:11.1570 &  -155.2828 &    19.4123 &  -0.270 & 1 & 5.40 & 5     &    12 &     5 & 216 &   7 &    8 & 84 \\
 6 & 2018-05-30T20:53:50.0830 &  -155.2852 &    19.4113 &  -0.070 & 1 & 5.40 & 2     &    19 &   -23 & 163 &  11 &  -43 & 87 \\
 7 & 2018-06-01T23:37:15.5938 &  -155.2765 &    19.3988 &   0.421 & 1 & 5.40 & 2     &    18 &   -15 & 201 &  19 &  -22 & 88 \\
 8 & 2018-06-04T01:50:49.9187 &  -155.2814 &    19.4122 &   1.168 & 1 & 5.40 & 2     &    21 &   -15 & 176 &   9 &  -31 & 89 \\
 9 & 2018-06-05T14:32:34.4450 &  -155.2860 &    19.4008 &  -0.830 & 1 & 5.50 & 2     &    21 &   -35 & 183 &  23 &  -46 & 87 \\
10 & 2018-06-07T02:06:38.8952 &  -155.2802 &    19.4113 &   1.362 & 1 & 5.40 & 2     &    21 &   -16 & 191 &  16 &  -28 & 87 \\
11 & 2018-06-08T12:44:39.9650 &  -155.2827 &    19.4138 &  -0.970 & 1 & 5.30 & 1     &    25 &   -32 & 137 &   9 &  -73 & 87 \\
12 & 2018-06-09T14:48:18.8490 &  -155.2805 &    19.4129 &   0.804 & 1 & 5.40 & 2     &    25 &   -31 & 172 &  16 &  -50 & 88 \\
13 & 2018-06-10T10:51:02.2379 &  -155.2805 &    19.4132 &   0.943 & 1 & 5.40 & 2     &    25 &   -31 & 172 &  16 &  -50 & 88 \\
14 & 2018-06-11T14:43:54.4867 &  -155.2829 &    19.4083 &   0.503 & 1 & 5.40 & 2     &    26 &   -30 & 188 &  20 &  -44 & 88 \\
15 & 2018-06-12T11:52:51.1808 &  -155.2787 &    19.4145 &   1.289 & 1 & 5.50 & 2     &    25 &   -42 & 199 &  30 &  -51 & 88 \\
16 & 2018-06-13T13:39:37.7992 &  -155.2796 &    19.4158 &   0.915 & 1 & 5.40 & 2     &    28 &   -29 & 115 &   8 &  -90 & 87 \\
17 & 2018-06-14T13:19:37.7572 &  -155.2802 &    19.4139 &   1.079 & 1 & 5.50 & 2     &    24 &   -33 & 167 &  16 &  -57 & 87 \\
18 & 2018-06-15T21:56:40.0360 &  -155.2853 &    19.4115 &   1.258 & 1 & 5.50 & 2     &    25 &   -41 & 197 &  29 &  -49 & 86 \\
19 & 2018-06-16T20:18:18.8038 &  -155.2887 &    19.4127 &   1.549 & 2 & 5.50 & 2     &    29 &   -30 & 151 &  17 &  -67 & 87 \\
20 & 2018-06-17T16:26:13.3561 &  -155.2836 &    19.4148 &   1.701 & 2 & 5.60 & 2     &    29 &   -39 & 161 &  24 &  -70 & 85 \\
21 & 2018-06-18T16:12:49.9138 &  -155.2802 &    19.4128 &   1.474 & 1 & 5.40 & 2     &    25 &   -39 & 207 &  33 &  -45 & 86 \\
22 & 2018-06-19T15:05:34.4552 &  -155.2840 &    19.4154 &   1.784 & 2 & 5.30 & 1     &    13 &   -34 & 224 &  41 &  -59 & 87 \\
23 & 2018-06-20T14:22:23.3430 &  -155.2862 &    19.4143 &  -0.020 & 1 & 5.40 & 2     &    26 &   -40 & 245 &  58 &  -60 & 85 \\
24 & 2018-06-21T23:12:59.9939 &  -155.2793 &    19.4085 &   0.210 & 1 & 5.40 & 2     &    25 &   -41 & 197 &  29 &  -49 & 83 \\
25 & 2018-06-23T04:52:16.6877 &  -155.2715 &    19.4020 &   0.837 & 1 & 5.20 & 1     &    24 &   -34 & 227 &  43 &  -48 & 82 \\
26 & 2018-06-24T02:34:37.7015 &  -155.2777 &    19.4131 &   1.131 & 1 & 5.20 & 1     &    28 &   -39 & 210 &  35 &  -45 & 82 \\
27 & 2018-06-25T02:12:24.4592 &  -155.2793 &    19.4177 &   0.589 & 1 & 5.50 & 2     &    16 &   -58 & 229 &  40 &  -88 & 85 \\
28 & 2018-06-26T03:02:45.5098 &  -155.2836 &    19.4113 &   1.655 & 2 & 5.30 & 1     &     4 &   -44 &  50 &  55 &  -83 & 86 \\
29 & 2018-06-27T08:40:40.0691 &  -155.2836 &    19.4113 &   1.700 & 2 & 5.60 & 2     &    18 &   -50 &  51 &  55 &  -87 & 87 \\
30 & 2018-06-28T14:48:50.0711 &  -155.2829 &    19.4159 &   1.162 & 1 & 5.60 & 2     &    22 &   -57 & 239 &  51 &  -76 & 84 \\
31 & 2018-06-29T17:50:46.6834 &  -155.2858 &    19.4137 &   0.974 & 1 & 5.60 & 2     &    22 &   -57 & 239 &  51 &  -76 & 84 \\
32 & 2018-07-01T00:51:13.3348 &  -155.2838 &    19.4140 &   1.355 & 1 & 5.50 & 2     &    13 &   -56 & 225 &  38 &  -87 & 83 \\
33 & 2018-07-02T11:24:46.6190 &  -155.2927 &    19.4032 &  -0.360 & 1 & 5.60 & 2     &    21 &   -56 &  47 &  53 &  -89 & 81 \\
34 & 2018-07-03T12:17:03.3810 &  -155.2747 &    19.4160 &  -0.120 & 1 & 5.50 & 2     &    16 &   -58 &  53 &  51 &  -81 & 83 \\
35 & 2018-07-04T20:19:11.1204 &  -155.2729 &    19.4038 &   0.575 & 1 & 5.50 & 2     &    16 &   -52 &  70 &  60 &  -71 & 81 \\
36 & 2018-07-05T23:20:04.4998 &  -155.2816 &    19.4077 &   0.905 & 1 & 5.50 & 2     &    13 &   -56 & 225 &  38 &  -87 & 82 \\
37 & 2018-07-07T04:04:38.8621 &  -155.2785 &    19.4012 &   0.409 & 1 & 5.60 & 2     &    16 &   -58 &  53 &  51 &  -81 & 83 \\
38 & 2018-07-08T12:54:50.0330 &  -155.2912 &    19.4020 &  -1.010 & 1 & 5.60 & 2     &    17 &   -57 &  55 &  54 &  -78 & 84 \\
39 & 2018-07-09T19:20:46.6280 &  -155.2897 &    19.4002 &  -0.470 & 1 & 5.70 & 2     &    22 &   -57 & 239 &  51 &  -76 & 84 \\
40 & 2018-07-11T15:45:53.3220 &  -155.2887 &    19.3993 &  -0.500 & 1 & 5.60 & 2     &    19 &   -53 &  70 &  58 &  -71 & 83 \\
41 & 2018-07-13T00:42:27.7436 &  -155.2790 &    19.4023 &   1.155 & 1 & 5.50 & 2     &    15 &   -53 &  58 &  60 &  -75 & 81 \\
42 & 2018-07-14T05:08:03.3680 &  -155.2768 &    19.3995 &  -0.210 & 1 & 5.60 & 2     &    17 &   -57 &  55 &  54 &  -78 & 83 \\
43 & 2018-07-15T13:26:05.5039 &  -155.2804 &    19.4079 &   0.845 & 1 & 5.60 & 2     &    16 &   -55 &  56 &  53 &  -86 & 82 \\
44 & 2018-07-16T21:42:36.6299 &  -155.2727 &    19.4017 &   0.641 & 1 & 5.50 & 2     &    15 &   -53 &  58 &  60 &  -75 & 83 \\
45 & 2018-07-18T11:28:04.4791 &  -155.2734 &    19.4020 &   1.000 & 1 & 5.60 & 2     &    17 &   -57 &  55 &  54 &  -78 & 84 \\
46 & 2018-07-20T02:33:02.2956 &  -155.2782 &    19.4046 &   1.077 & 1 & 5.60 & 2     &    16 &   -58 &  53 &  51 &  -81 & 84 \\
47 & 2018-07-21T19:43:29.9534 &  -155.2770 &    19.4038 &   0.601 & 1 & 5.60 & 2     &    16 &   -58 &  53 &  51 &  -81 & 81 \\
48 & 2018-07-23T06:53:39.9521 &  -155.2760 &    19.4051 &   0.633 & 1 & 5.60 & 2     &    16 &   -58 &  53 &  51 &  -81 & 84 \\
49 & 2018-07-24T16:41:10.0831 &  -155.2807 &    19.4104 &   1.070 & 1 & 5.60 & 2     &    19 &   -53 &  70 &  58 &  -71 & 83 \\
50 & 2018-07-26T22:09:11.1978 &  -155.2751 &    19.4087 &   1.114 & 1 & 5.60 & 2     &    16 &   -55 &  56 &  53 &  -86 & 82 \\
51 & 2018-07-28T12:37:26.6136 &  -155.2692 &    19.4039 &   0.377 & 1 & 5.60 & 2     &    16 &   -58 &  53 &  51 &  -81 & 78 \\
52 & 2018-07-29T22:10:26.6263 &  -155.2815 &    19.4083 &   1.543 & 2 & 5.40 & 2     &    10 &   -46 &  52 &  53 &  -77 & 83 \\
53 & 2018-07-31T17:59:47.7051 &  -155.2742 &    19.4104 &   1.322 & 1 & 5.50 & 2     &    16 &   -52 &  70 &  60 &  -71 & 83 \\
54 & 2018-08-02T21:55:13.3274 &  -155.2783 &    19.4107 &   1.451 & 1 & 5.60 & 2     &    17 &   -57 &  55 &  54 &  -78 & 86 \\
    \end{tabular}
\end{sidewaystable}
\clearpage\pagebreak

\section*{Acknowledgements}
We thank Jo{\"e}l Ruch for illuminating discussions about faulting and collapse styles at \kila\ and other calderas,
and Walter Tape \& Carl Tape for discussion about moment tensors.
R.S.M. was supported by NSF grant EAR-1446543.
\def \agu{Am.~Geophys.~Un.} \def \acm{Comm. Assoc. Computing Machinery} \def
  \ess{Earth and Space~Sci.} \def \gres{Gondwana Research} \def
  \jgrpl{J.~Geophys.~Res. Planets} \def \jrmms{J.~Rock~Mech. \& Mining~Sci.}
  \def \landsl{Landslides} \def \lpsc{Proc. Lunar Planet. Sci. Conf.}\def
  \natcom{Nature~Communications} \def \pass{Planet. Space~Sci.} \def
  \pre{Phys.~Rev.~E} \def \prl{Phys.~Rev.~Lett.} \def \pr{Physics~Reports} \def
  \sjg{Swiss~J.~Geosci.} \def \ssr{Space Sci. Rev.} \def \bell{Bell~Sys.
  Tech.~J.}\def \pire{Proc.~Inst.~Rad.~Eng.}\def \jsp{J.~Stat.~Phys.}\def
  \agu{Am.~Geophys.~Un.} \def \usgs{U.S.~Geol.~Survey} \def \dggs{Alaska Div.
  Geol. Geophys. Surv.} \def \antsci{Antarctic~Science} \def \aapg{Am.~Assoc.
  Petroleum~Geol.} \def \aapgb{\aapg~Bull.} \def \aapgm{\aapg~Memoir} \def
  \acha{Applied Comput. Harmonic Analysis} \def \actag{Acta~Geophysica} \def
  \agt{Acta Geologica Taiwanica} \def \amsci{Am.~Sci.} \def \ajs{Am.~J.~Sci.}
  \def \angeo{Annals.~Geophy.} \def \areps{Annu.~Rev. Earth Planet.~Sci.} \def
  \aa{Astron.~Astophys.} \def \ag{Astron.~Geophys.} \def \aj{Astrophys.~J.}
  \def \araa{Annu.~Rev. Astron.~Astrophys.} \def \bssa{Bull.~Seismol.~Soc.~Am.}
  \def \basinr{Basin~Research} \def \bvolc{Bull.~Volcanology} \def
  \beri{Bull.~Earthquake Research~Inst.} \def \biesas{Bull. Inst. Earth Sci.,
  Academia Sinica} \def \ccp{Commun. Comput. Phys.} \def \cse{Computing
  Science~Engineering} \def \csd{Computational Science \& Discovery} \def
  \cg{Computers \& Geosciences} \def \comphys{Computers in Physics} \def
  \chemgeo{Chem.~Geol.} \def \cj{Computer Journal} \def \cjes{Can.~J.
  Earth~Sci.} \def \cnsns{Comm.~Nonlin.~Sci. Num.~Sim.} \def \crst{Cold~Regions
  Sci.~Tech.} \def \dao{Dynamics~Atmos.~Oceans} \def \ecgeo{Econ.~Geol.} \def
  \eqspec{Earthquake Spectra} \def \eps{Earth~Planets~Space} \def
  \esr{Earth-Sci.~Rev.} \def \epsl{Earth~Planet. Sci.~Lett.} \def
  \eos{Eos~Trans. \agu} \def \fb{First~Break} \def \geol{Geology} \def
  \gsa{Geol.~Soc.~Am.} \def \gsat{GSA~Today} \def \gsab{Geol.~Soc.~Am. Bull.}
  \def \geomag{Geol.~Mag.} \def \geop{Geophysics} \def \geos{Geosphere} \def
  \ggg{Geochem.~Geophy.~Geosyst.} \def \gi{Geof\'isica.~Internacional} \def
  \geophysj{Geophys.~J.} \def \gji{Geophys.~J.~Int.} \def
  \grl{Geophys.~Res.~Lett.} \def \gjras{Geophys.~J. R.~Astron.~Soc.} \def
  \gml{Geo-Marine~Lett.} \def \gms{Geophys.~Monogr.~Series} \def
  \gp{Geophys.~Prosp.} \def \ia{Island~Arc} \def
  \ieee{IEEE~Trans.~Vis.~Comp.~Graphics} \def \igr{International Geology
  Review} \def \ijnme{Int. J. Numerical Meth. Engineering} \def
  \ip{Inverse~Problems} \def \jam{J.~App.~Mech.} \def \jaes{J.~Asian
  Earth~Sci.} \def \jasa{J.~Acoust.~Soc.~Am.} \def \jastp{J.~Atmos.
  Solar-Terr.~Phys.} \def \jclim{J.~Climate} \def \jem{J.~Eng.~Mech.} \def
  \jcp{J.~Comp.~Phys.} \def \jfm{J.~Fluid~Mech.} \def \jgr{J.~Geophys.~Res.}
  \def \jgrse{J.~Geophys.~Res. Solid Earth} \def \jg{J.~Geodynamics} \def
  \jgeo{J.~Geology} \def \jgeop{J.~Geophysics} \def \jgsl{J.~Geol.~Soc. London}
  \def \jseis{J.~Seis.} \def \jmg{J.~Metamorphic~Geol.} \def
  \jmp{J.~Math.~Phys.} \def \jnaiam{J.~Num.~Analysis, Industrial App. Math.}
  \def \jota{J.~Optim. Th.~App.} \def \jpe{J.~Phys.~Earth} \def
  \jrssb{J.~R.~Statist.~Soc.~B} \def \jsaes{J.~South American Earth Sciences}
  \def \jsc{J.~Sci.~Comput.} \def \jsg{J.~Struct.~Geol.} \def
  \jsr{J.~Sed.~Res.} \def \jscs{J.~Statist. Comput.~Simul.} \def
  \jvgr{J.~Volcan. Geothermal~Res.} \def \ledge{Leading~Edge} \def
  \lncs{Lecture~Notes in Computer~Science} \def \lith{Lithosphere} \def
  \mg{Marine~Geology} \def \mgr{Marine Geophysical Researches} \def
  \mathgeo{Mathematical Geology} \def \mgsc{Mem. Geol. Soc. China} \def
  \mi{Math.~Intelligencer} \def \mnras{Mon.~Not. R.~Astron.~Soc.} \def
  \mpg{Marine Petroleum Geology} \def \mwr{Monthly Weather Review} \def
  \nat{Nature} \def \natgeo{Nature~Geoscience} \def \natphys{Nature~Physics}
  \def \nathaz{Natural~Hazards} \def \nzjgg{New.~Zealand J.~Geol.~Geophys.}
  \def \numa{Numerical Algorithms} \def \ps{Polar~Science} \def
  \pt{Physics~Today} \def \pce{Physics and Chemistry of the Earth} \def
  \pepi{Phys.~Earth Planet.~Inter.} \def \ptrsl{Phil.~Trans. R.~Soc.~Lond.}
  \def \ptrsA{Phil.~Trans. R.~Soc.~A.} \def \ptrsla{Phil.~Trans.
  R.~Soc.~Lond.~A.} \def \pag{Pure~App.~Geophys.} \def \pnas{Proc.~Natl.
  Acad.~Sci.} \def \prsa{Proc.~R.~Soc.A} \def \pieee{Proc.~IEEE} \def
  \pers{Photogrammetric Eng. \& Remote Sensing} \def
  \qjrms{Q.~J.~R.~Meteorol.~Soc.} \def \qsr{Quaternary Sci.~Rev.} \def
  \rpp{Rep.~Prog.~Phys.} \def \rgsp{Rev.~Geophys.~Space.~Phys.} \def
  \rgp{Rev.~Geophys.} \def \rmp{Rev.~Mod.~Phys.} \def \sa{Sci.~Am.} \def
  \se{Solid~Earth} \def \sci{Science} \def \scirep{Sci.~Rep.} \def
  \scipro{Science~Progress} \def \sciadv{Science~Advances} \def
  \srl{Seismol.~Res.~Lett.} \def \spej{Soc.~Petroleum Engineers~J.} \def
  \sp{Solar~Physics} \def \segea{SEG Expanded Abstracts} \def \seg{Soc.
  Economic Geologists} \def \sepm{Soc. Sedimentary Geology} \def
  \sirev{SIAM~Rev.} \def \sjna{SIAM~J. Numer.~Anal.} \def \sjsc{SIAM~J.
  Sci.~Comp.} \def \sjssc{SIAM~J. Sci.~Stat.~Comp.} \def
  \statsci{Statistical~Science} \def \survgp{Surv.~Geophys.} \def
  \tao{Terr.~Atmos. Oceanic~Sci.} \def \tec{Tectonics} \def \ternov{Terra~Nova}
  \def \tecphy{Tectonophysics} \def \wm{Wave~Motion} \def \zis{Zisin (J.~Seis.
  Soc.~Japan)}

\end{document}